\documentclass[11pt,english,aps,manuscript]{revtex4}
\usepackage{graphicx}
\usepackage{dcolumn}
\usepackage{array}
\usepackage{amsmath}
\usepackage{amsfonts}

\usepackage[utf8]{inputenc}

\usepackage{braket}
\usepackage{tabularx}

\usepackage{mathrsfs}
\usepackage{babel}
\usepackage{textcomp}

\newcolumntype{Y}{>{\centering\arraybackslash}X}

\newcommand{\wigmtx}[4]{\mathcal{D}_{#2 \, #3}^{\,#1}\left(#4\right)}

\begin{document}


\title{Prospects for Quantum Computing with an Array \\ of Ultracold Polar Paramagnetic Molecules}

\makeatother

\author{Mallikarjun Karra, Ketan Sharma and Bretislav Friedrich}

\thanks{Author to whom correspondence should be addressed. Electronic mail: bretislav.friedrich@fhi-berlin.mpg.de}

\affiliation{Fritz-Haber-Institut der Max-Planck-Gesellschaft, Faradayweg 4-6,
D-14195 Berlin, Germany}

\author{Sabre Kais}

\affiliation{Departments of Chemistry, Physics  and Birck Nanotechnology Center, Purdue University,
West Lafayette, IN 47907, U.S.A.}

\author{Dudley Herschbach}

\affiliation{Department of Chemistry and Chemical Biology, Harvard University, 12 Oxford Street, Cambridge, MA 02138, U.S.A.}

\date{\today}

\begin{abstract}

Arrays of trapped ultracold molecules represent a promising platform for implementing a universal quantum computer. DeMille [\textit{Phys. Rev. Lett.}, \textbf{88}, 067901 (2002)] has detailed a prototype design based on Stark states of polar $^1\Sigma$ molecules as qubits. Herein, we consider an array of polar $^2\Sigma$ molecules which are, in addition, inherently paramagnetic and whose Hund's case (b) free-rotor states are Bell states. We show that by subjecting the array to combinations of concurrent homogeneous and inhomogeneous electric and magnetic fields, the entanglement of the array's Stark and Zeeman states can be tuned and the qubit sites addressed. Two schemes for implementing an optically controlled CNOT gate are proposed and their feasibility discussed in the face of the broadening of spectral lines due to dipole-dipole coupling and the inhomogeneity of the electric and magnetic fields.

\vspace{5 cm}
\end{abstract}

\maketitle


\section{Introduction}
\label{Introduction}

Since its inception in 1982 by Feynman \cite{Feynman} and follow-up work by others \citep{EarlyQCpapers1, Deutsch97_EarlyQCpapers2, EarlyQCpapers3}, the idea of a universal quantum computer has been pursued and amplified in many quarters. Whether for reasons of fundamental interest \citep{Ansmann2009,Weber2014,RevModPhys.86.153} or because of the promise of a computational advantage \citep{QC_Review1,QC_Review2,ShorAlgo,GroverAlgo,Sabre_Advances_2014}, these pursuits identified a number of physical systems \citep{Proposals1, Proposals2, Proposals3, Proposals4, Proposals5, Proposals6,NielsonBook1,QuantumCircuitModel,One-way_QC} that meet the DiVincenzo requirements \citep{DiVincenzoCriteria} for the physical implementation of quantum computation \citep{NaturePhysicsQuantumSimul}. 

Among the candidate systems  has been an array of optically trapped ultra-cold polar molecules, first proposed and investigated by DeMille \citep{DeMille}. This seminal work demonstrated how dipole-dipole interactions between polar $^1\Sigma$ molecules trapped in a one-dimensional optical array would allow fast and efficient quantum control by  resonant laser drive pulses with little decoherence. Our subsequent work examined aspects of DeMille's proposal for polar closed-shell molecules, whether linear \citep{Trinity1,Trinity3} or symmetric tops \citep{Trinity2}. 

Herein, we consider an array of trapped ultra-cold polar $^2\Sigma$ molecules that are open-shell and whose nonzero electronic spin makes them inherently paramagnetic. An array of such molecules is entangled  by the electric dipole-dipole interaction and subject to  combinations of concurrent homogeneous and inhomogeneous electric and magnetic fields. Since a sequence of single qubit gates and CNOT gates is sufficient to build a unitary-evolution based universal quantum computer \citep{TwoBitUQC}, our  objective is to assess the feasibility of implementing a CNOT logic gate, i.e., a gate that flips the target qubit depending on the state of the control qubit. 

We characterized the eigenstates of the array, including their mutually induced directionality, by evaluating their eigenproperties via numerical diagonalization of the appropriate Hamiltonian matrix, whose elements we found analytically. We also evaluated the concurrence of the states as a measure of their entanglement in the presence and absence of fields. A key feature of the system is that in the absence of fields, its states are all the maximally entangled Bell states. Applying an inhomogeneous magnetic field disentangles these states and can be used to perform a Bell measurement. This feature may be of consequence for superdense coding \citep{PhysRevLett.69.2881} and quantum teleportation \citep{PhysRevLett.70.1895}. 	
	
Our findings led us to propose two novel schemes for implementing an optically controlled CNOT gate operation. Both  schemes make use of the adiabatic theorem and can be classified as adiabatic quantum computation \citep{AQC1, AQC2} (even though one of the three steps in both schemes is not adiabatic). 

Of key importance is the ability to resolve the transition frequencies involved in the optical control of the gate operations -- in the face of the broadening due to dipole-dipole coupling and the inhomogeneity of the electric and magnetic fields. We  show that the former dominates over the latter and set the criteria for the feasibility of the schemes. 

The paper is organised as follows: In Section \ref{Sec: 2}, we briefly discuss the Hamiltonian of a system of two $^2\Sigma$ molecules interacting via electric dipole-dipole interaction in the presence of concurrent electric and magnetic fields. We then describe our choice of qubits and explain the behavior of the lowest four $\tilde{N}=0$ eigenstates of the system before illustrating the proposed schemes for quantum CNOT logic gate implementation in Section \ref{Sec: 3}. Key results for a pair of NaO molecules as a model system are presented in Section \ref{Sec: 4}, wherein we also revisit the issue of broadening of spectral lines due to dipole-dipole coupling and inhomogeneity of the field(s) at the two qubit sites. Finally, in Section \ref{Sec: 5} we summarize our  results and offer  prospects for extensions and applications of the work done so far.

\section{A unit quantum circuit: A pair of $^2\Sigma$ molecules}
\label{Sec: 2}
\subsection{Hamiltonian}

\begin{figure}
\centering
\includegraphics[width=01\textwidth, height=\textheight, keepaspectratio]{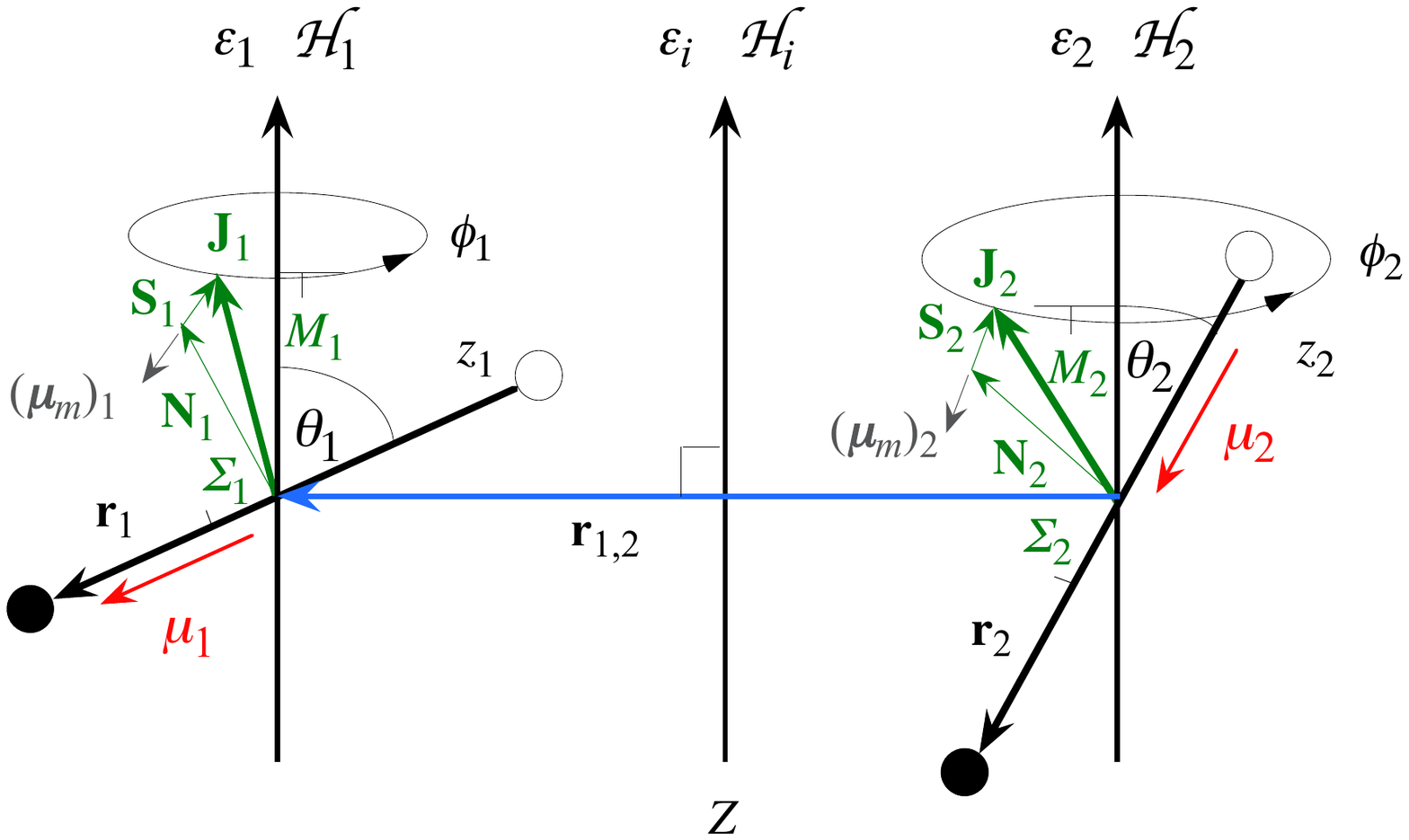}
\caption{The configuration for a system of two polar $^2\Sigma$ molecules in an optical array with superimposed (in)homogeneous and concurrent electric and/or magnetic fields. See text.}
\label{fig:ModelSystem}
\end{figure}

The Hamiltonian of a pair of $^2\Sigma$ molecules in the presence of concurrent electric and magnetic fields is the sum of the single-molecule  Hamiltonians, $H_i$, and the electric and magnetic dipole-dipole coupling terms. Upon neglecting the much weaker magnetic dipole-dipole interaction, the Hamiltonian takes the form
\begin{equation}
\label{eqn:Main}
H = \sum_{i=1}^2H_{i} + V_{d-d},
\end{equation}
where  $i=1,2$ and $V_{d-d}$ is the electric dipole-dipole interaction.

The single-molecule Hamiltonian (apart from nuclear spin) is given by the sum of the rotational, Stark and Zeeman terms \citep{KetanP1, StericProficiency}.  
\begin{equation}
H_{i}=B_{i}\textbf{N}_{i}^{2} + \gamma_{i}\textbf{N}_{i} \cdot \textbf{S}_{i} - B_{i}(\eta_{el})_i\cos{\boldsymbol{\theta}_{i}} + B_{i}(\eta_{m})_i(\textbf{S}_{Z})_{i}
\label{eqn:Hami}
\end{equation}
where $B_i$ is the rotational constant, $\textbf{N}_i$ the rotational angular momentum operator, $\textbf{S}_i$ the electronic spin angular momentum operator, $\gamma_i$  the spin-rotation coupling constant and $(\textbf{S}_Z)_i$ the space-fixed $Z$ component of the electronic spin of the $i$-th molecule. The dimensionless magnetic and electric interaction parameter of the $i$-th molecule is given respectively by
\begin{equation}
\eta_m\equiv\frac{\mu_m \mathcal{H}}{B}
\label{etam}
\end{equation}
and
\begin{equation}
\eta_{el}\equiv\frac{\mu\varepsilon}{B}
\label{etael}
\end{equation}
where $\mu_m=g_S\mu_B$ is the electronic magnetic dipole moment of the $^2\Sigma$ molecule, $g_S \cong 2.0023$ the electronic gyromagnetic ratio, $\mu_B$ the Bohr magneton, $\mu$ the body-fixed electric dipole moment, and $\mathcal{H}$ the magnetic and $\varepsilon$ the electric field strength. 

The magnetic and electric fields $\mathcal{H}$ and $\varepsilon$ are assumed to be collinear and their common direction defines the space-fixed axis $Z$, see Figure \ref{fig:ModelSystem}. The  electric dipole-dipole interaction potential is given by \cite{Trinity1}
\begin{equation}
V_{d-d}=\frac{\boldsymbol{\mu}_1 \cdot \boldsymbol{\mu}_2-3(\boldsymbol{\mu}_1 \cdot {\bf n})(\boldsymbol{\mu}_2 \cdot {\bf n})}{r_{1,2}^3}
\label{Vdd}
\end{equation}
with $\boldsymbol{\mu}_1$ and $\boldsymbol{\mu}_2$ the electric dipole moments of the two molecules  and $\bold{r}_{1,2}$ the relative position vector of the centres of mass of the two molecules whose direction is given by the unit vector $\bold{n}\equiv \frac{\bold{r}_{1,2}}{r_{1,2}}$. As usual, $r_{1,2} \equiv |{\bf r_{1,2}}|$ and $\mu_{1,2} \equiv |\boldsymbol{\mu}_{1,2}|$. Moreover, in our case, $\mu_{1}=\mu_{2}\equiv \mu$. 

Eq. (\ref{eqn:Xi}) can be recast  in terms of the Wigner matrices $\wigmtx{l}{m}{0}{\phi, \theta, \chi}$ \citep{KetanP2}:
\begin{equation}
\label{eqn:Xi}
 V_{d-d}=-\sqrt{6}~\Xi \sum_{\nu\,\lambda}C(1,1,2;\nu,\lambda,\nu+\lambda)\wigmtx{1}{-\nu}{0}{\phi_1,\theta_1,\chi_1}\wigmtx{1}{-\lambda}{0}{\phi_2,\theta_2,\chi_2}\wigmtx{2}{\nu+\lambda}{0}{\phi,\theta,\chi}
\end{equation}
where $C(J_1,J_2,J_3;M_1,M_2,M_3)$ are the Clebsch-Gordan coeffcients, $J_1$ and $J_2$ the angular momentum qunatum numbers of molecules 1 and 2, $M_1$ and $M_2$ the projection of the angular momenta of molecules 1 and 2 on the space fixed axis $Z$, $(\theta_1,\phi_1)$ and $(\theta_2,\phi_2)$ the rotational coordinates of molecules 1 and 2,  $(\theta,\phi)$ the spherical coordinates of their relative position vector ${\bf r_{1,2}}$, and $\Xi\equiv\mu_1\mu_2/r_{1,2}^3$ is a parameter that characterises the strength of the electric dipole-dipole interaction.

\begin{figure}
\centering
\includegraphics[width=1\textwidth, height=\textheight, keepaspectratio]{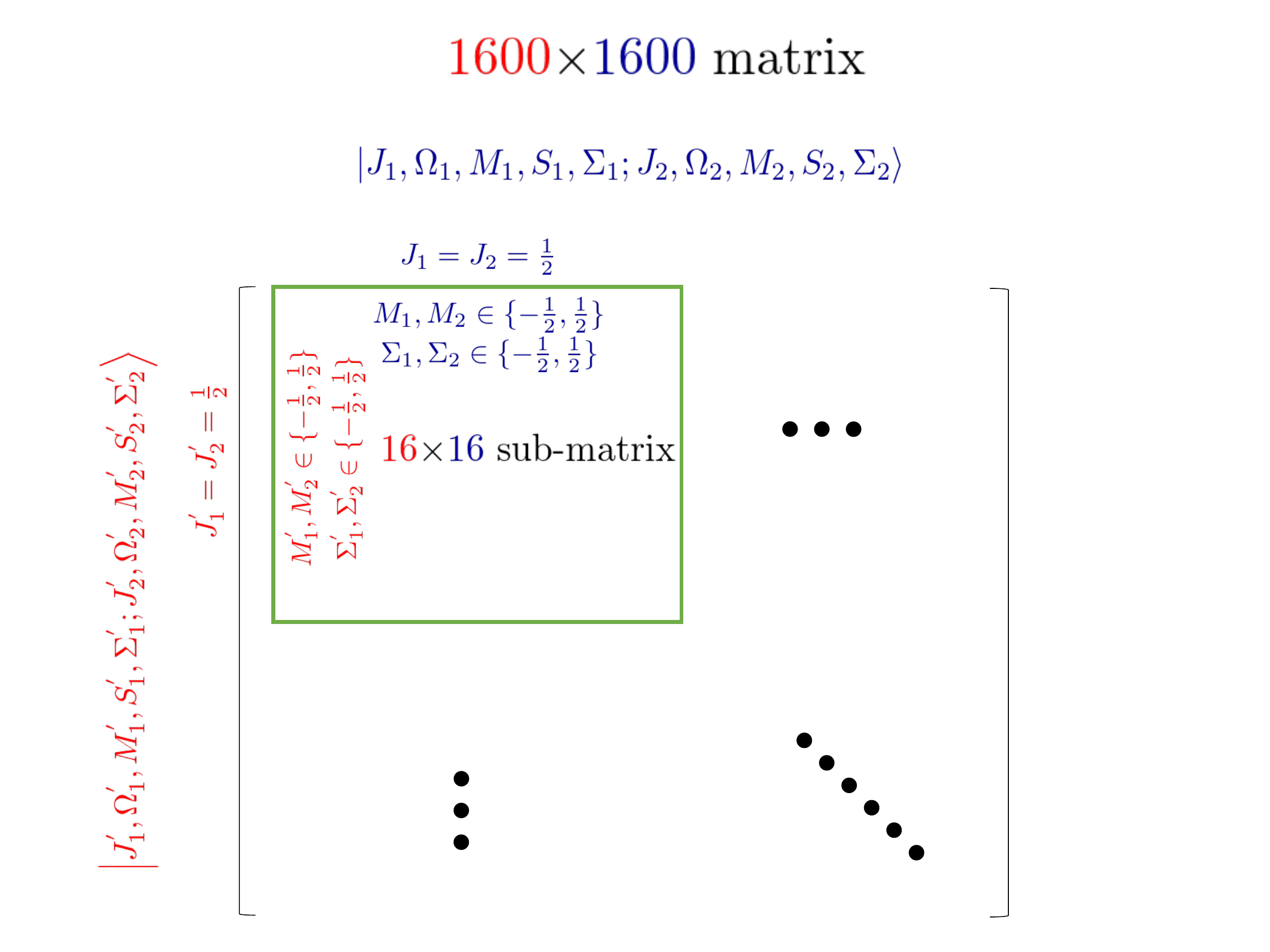}
\caption{Matrix representation of Hamiltonian of Eq. (\ref{eqn:Main}) in the cross product basis set $|J_1,\Omega_1,M_1,S_1,\Sigma_1;J_2,\Omega_2,M_2,S_2,\Sigma_2\rangle$ of two Hund's case (b) molecules, truncated such that $J_i$ with $i=1,2$ ranges from $\frac{1}{2}$ to $\frac{7}{2}$ for molecules $1$ and $2$. Hence $M_i$ ranges from $-J_i$ to $J_i$ while  $\Sigma_i=\pm\frac{1}{2}$. Same applies for primed quantities. Note that $J_1=J_2=\frac{1}{2}=J_1^{'}=J_2^{'}$ give rise to a $16 \times 16$ sub-matrix; the bottom four of the 16 states obtained by its diagonalization are the maximally entangled Bell states for our choice of qubits. See text.}
\label{fig:TruncatedHamiltonianSchematic}
\end{figure}

The matrix elements of the Hamiltonian were calculated analytically in the cross product basis set, $|J_1,\Omega_1,M_1,S_1,\Sigma_1;J_2,\Omega_2,M_2,S_2,\Sigma_2\rangle$, of the two molecule \cite{KetanP2} and the eigenproperties of the composite two-molecule system obtained by a numerical diagonalization of  a truncated Hamiltonian matrix, whose structure is shown in Figure \ref{fig:TruncatedHamiltonianSchematic}. Note that the projection quantum numbers $\Omega_i$ and $\Sigma_i$ (with $i=1,2$) of the electronic angular momenta on the body-fixed axis of each $^2\Sigma$ molecule coincide, i.e., $\Omega_i=\Sigma_i$. The number of pairs of states determines the size of the basis set and is given by $[2\Sigma_{J_{\min}}^{J_{\max}} (2J+1)]^2$. For $J_{\min}=\frac{1}{2}$ and $J_{\max}=\frac{7}{2}$, this means that the truncated Hamiltonian matrix is of a $1600$ rank.

\subsection{Choice of qubits}
\label{Sec: 2B}

\begin{figure}
\centering
\includegraphics[width=1\textwidth, height=\textheight, keepaspectratio]{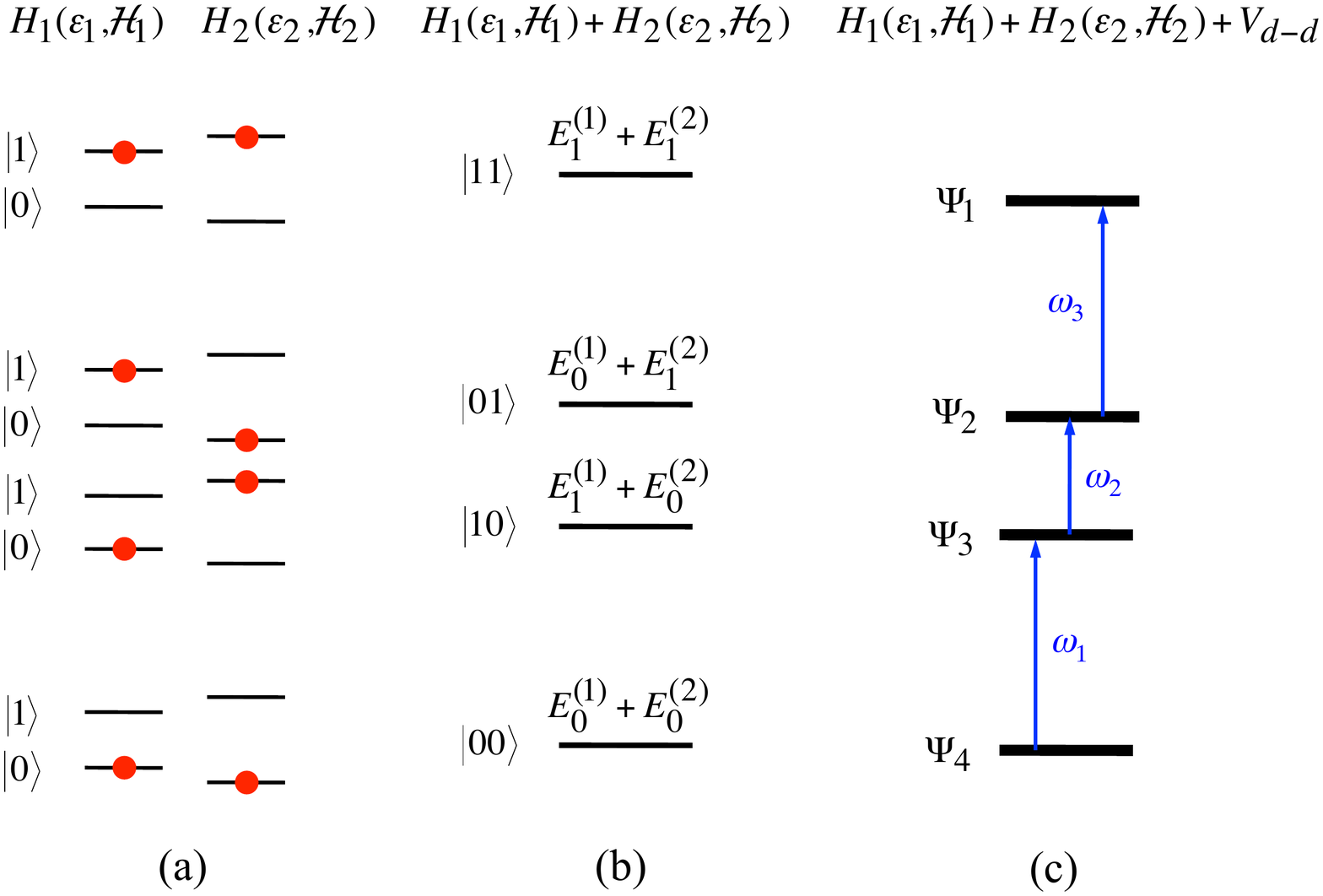}
\caption{A schematic of the  energy levels and basis states for a pair of molecules (labeled 1 and 2) in adjacent qubit sites (labeled 1 and 2) subject to electric and magnetic fields $\varepsilon$ and $\mathcal H$. States $|0\rangle$ and $|1\rangle$ with eigenenergies $E_0$ and $E_1$ of the individual molecules are chosen as qubits. (a) Levels of individual molecules; (b) Levels of the composite two-molecule system in the absence of dipole-dipole coupling; (c) Levels of the composite two-molecule system in the presence of dipole-dipole coupling. Also show are the transition frequencies between the states. See text.}
\label{fig:Schematic}
\end{figure}

\begin{figure}
\centering
\includegraphics[width=1\textwidth, height=\textheight, keepaspectratio]{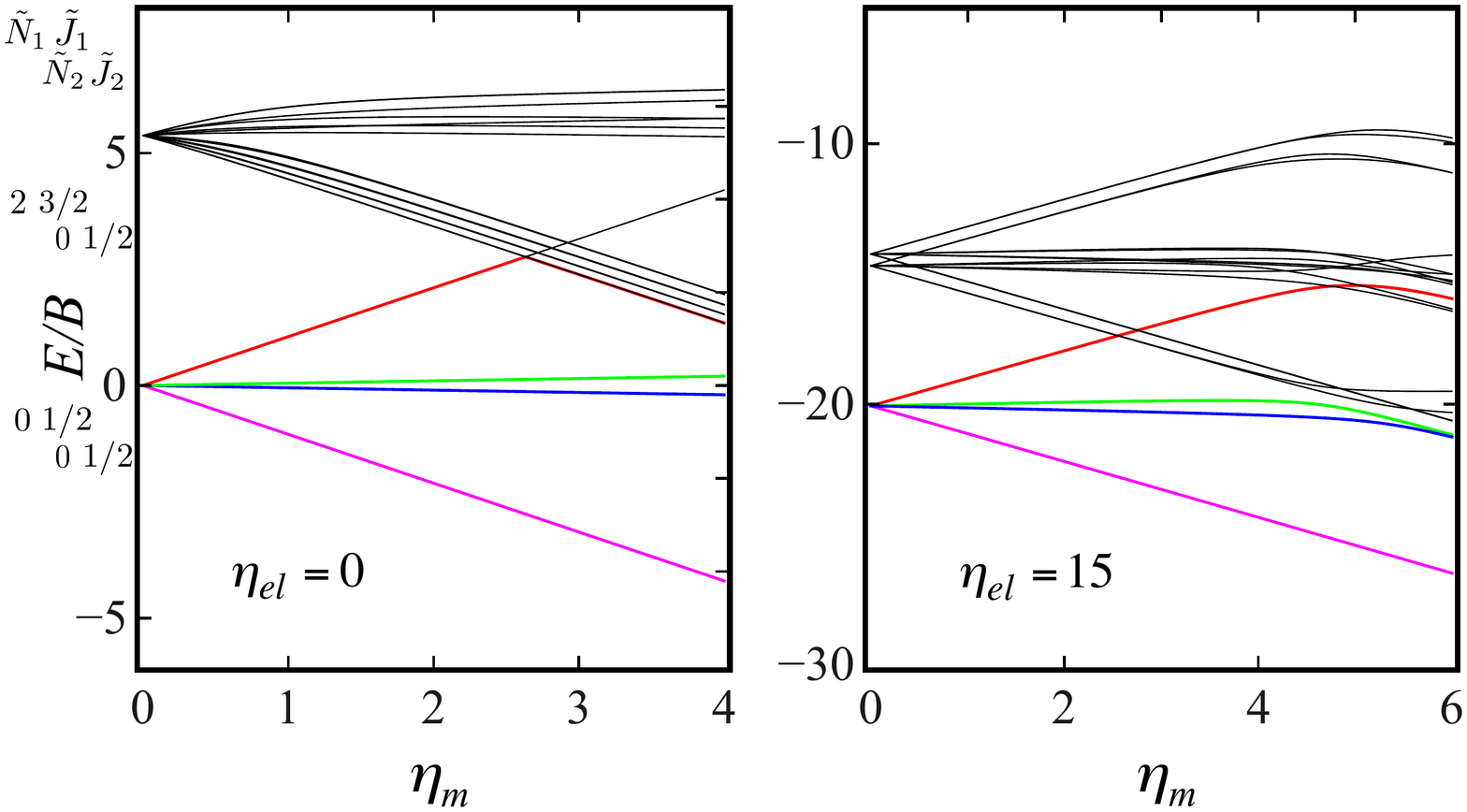}
\caption{Eigenenergies (in terms of the rotational constant $B$) of the composite two-molecule system with $\Xi/B\sim10^{-5}$ in an inhomogeneous magnetic field as functions of the magnetic field strength parameter $\eta_m$ for fixed values of the electric field strength parameter $\eta_{el}=0$ (left) and $\eta_{el}=15$. The inhomogeneity of the magnetic field is such that $(\eta_m)_1\equiv\eta_m=(\eta_m)_2/1.15$.}
\label{fig:AvoidedCrossings}
\end{figure}

A schematic of the energy levels and basis states for a pair of molecules in adjacent qubit sites is shown in Figure \ref{fig:Schematic} and the eigenenergies of the composite two-molecule system in an inhomogeneous magnetic field with and without an electric field are shown in Figure \ref{fig:AvoidedCrossings}. The top-most of the four $\tilde{N}=0$ states (red curve) exhibits an avoided crossing with one of the higher states for both  $\eta_{el}=0$ and $\eta_{el}=15$. 
Due to the opposite signs of the Stark and Zeeman terms in our Hamiltonian, Eq. (\ref{eqn:Hami}), a concurrent electric field can be used to tune the position -- and strength -- of such an avoided crossing, see also Refs \citep{PhysRevLett.97.083201,PhysRevA.80.033419}. Based on these results, we  chose the following states as qubits for the two CNOT schemes,
\begin{equation}
\label{eqn:basis}
\begin{aligned}
& \ket{0}=\Psi\left(\tilde{J}=\frac{1}{2}, \tilde{N}=0, M=-\frac{1}{2}\right)  \\
& \ket{1}=\Psi\left(\tilde{J}=\frac{1}{2}, \tilde{N}=0, M=+\frac{1}{2}\right)
\end{aligned}
\end{equation}
The field free quantum numbers $N$ and $J$ are no longer good quantum numbers in the field(s), but can be used as adiabatic labels, which is indicated by a tilde, $|\tilde{N},\tilde{J},M;\eta_{el},\eta_{m}\rightarrow 0\rangle \rightarrow |N,J,M\rangle$. The chief motivation for this choice of qubit states is that the field-free rotor states of a $^2\Sigma$ molecule -- which fall under Hund's case (b) \citep{StericProficiency,BobField_2015} -- are comprised of fully entangled combinations of such states. As we will see below, this offers some  advantages over the schemes presented in our previous work \citep{Trinity1,Trinity2,Trinity3}. In keeping with custom, we will refer to fully entangled combinations of qubit states as Bell states. 

\subsection{Behavior of a two-qubit system in concurrent electric and magnetic fields}

\begin{figure}
\centering
\includegraphics[width=1\textwidth, height=\textheight, keepaspectratio]{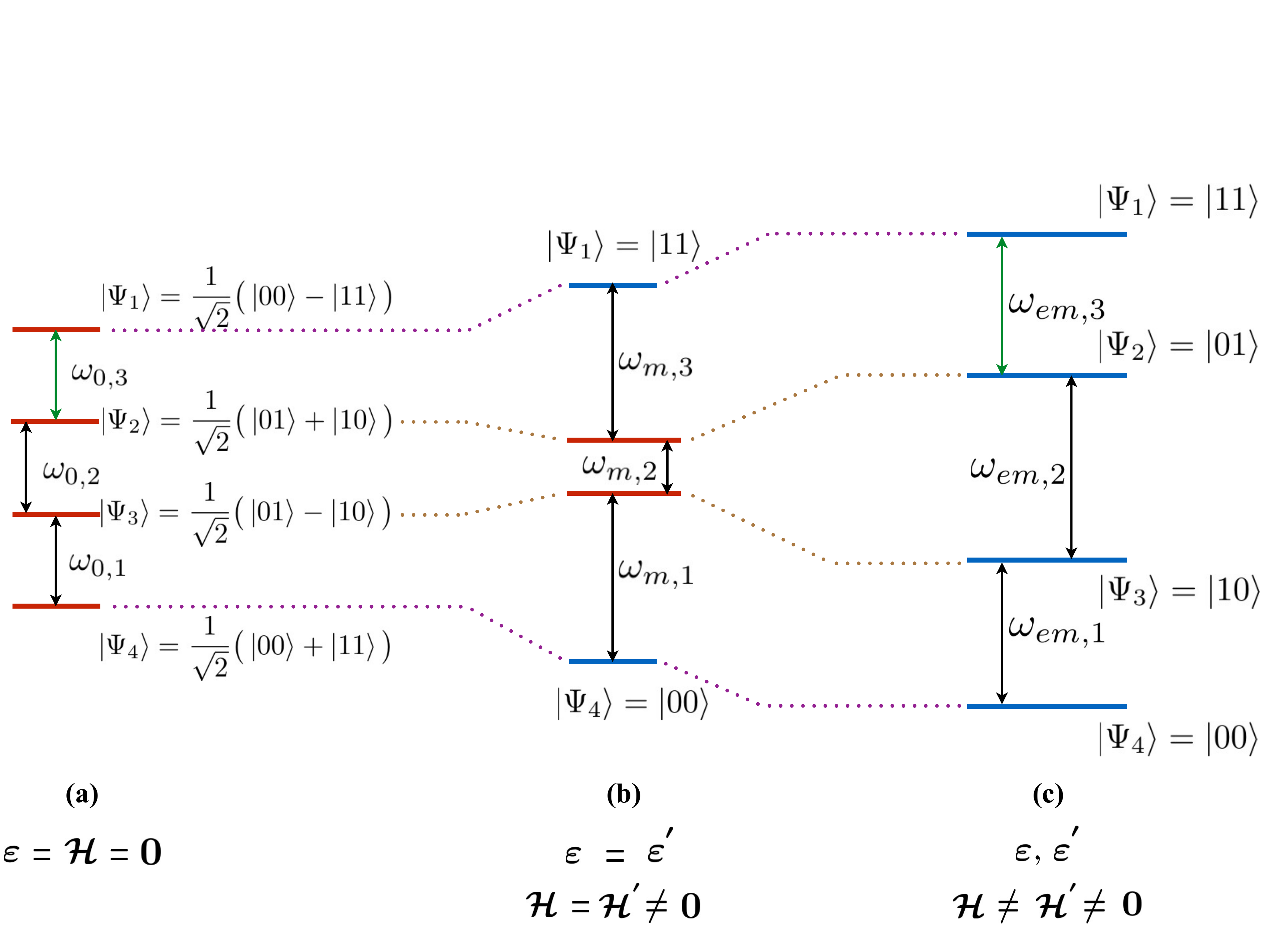}
\caption{Eigenenergies of the composite system of two $^2\Sigma$ molecules  coupled by the electric dipole-dipole interaction. The eigenstates are expressed in terms of the lowest four $\tilde{N}=0$ single-molecule eigenstates. In the absence of fields, the system can exist in any of the four Bell states, shown in red, panel (a). A homogeneous magnetic field is capable of disentangling the top and bottom states, leaving the two intermediate states entangled, as shown in panel (b). An inhomogeneous magnetic field disentangles the intermediate two states as well, panel (c). Frequencies important for effecting CNOT gate transitions are shown in green.}
\label{fig:EnergySchematic}
\end{figure}

The lowest four eigenstates of the composite system of two $^2\Sigma$ molecules can be written in terms of the  qubit states chosen above, Eq. (\ref{eqn:basis}). In the field-free case, these four eigenstates are the maximally entangled Bell states, cf. Fig. \ref{fig:TruncatedHamiltonianSchematic} and panel (a) of Fig. \ref{fig:EnergySchematic}. 

As shown in Fig. \ref{fig:EnergySchematic}, the Bell states are separated by the  frequencies $\omega_{0, 1}$, $\omega_{0, 2}$ and $\omega_{0, 3}$. When a homogeneous magnetic field is applied, the lowest and the highest state undergoes a disentanglement, while the states in between retain their maximal entanglement; the separation of the states is characterised by the frequencies $\omega_{m, 1}$, $\omega_{m, 2}$ and $\omega_{m, 3}$ , see panel (b) of Fig. \ref{fig:EnergySchematic}). A superimposed  inhomogeneous magnetic field differentiates between the two molecules and disentangles even the intermediate two states, see panel (c) of Fig. \ref{fig:EnergySchematic}). The resulting four states are separated by  frequencies $\omega_{em, 1}$, $\omega_{em, 2}$ and $\omega_{em, 3}$. We note that an electric field would couple more states, whereby many more avoided crossings would be generated, cf. panel (b) of Fig. \ref{fig:AvoidedCrossings}.

\begin{figure}
\centering
\includegraphics[width=0.81\textwidth, height=\textheight, keepaspectratio]{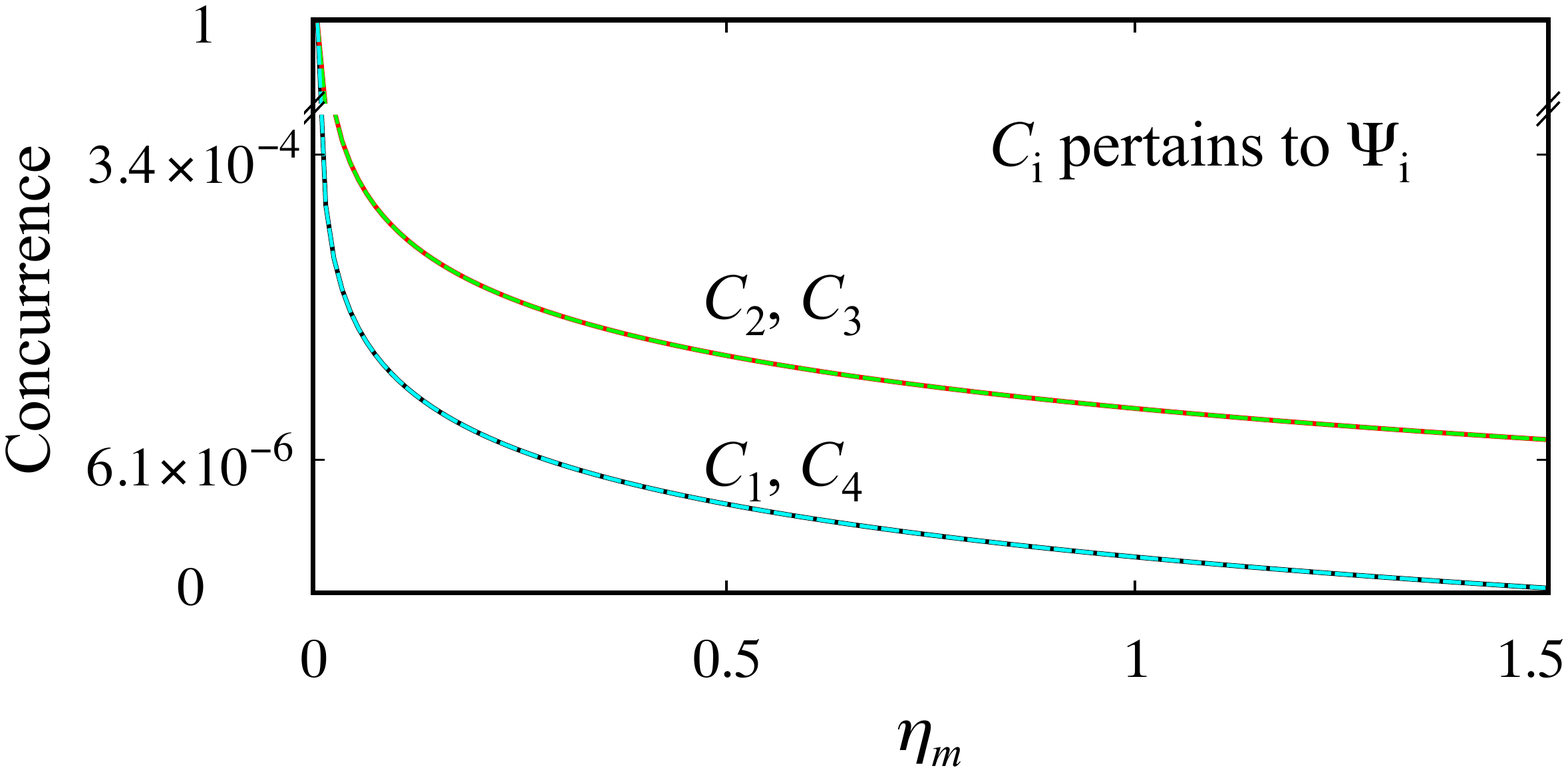}
\caption{Concurrences $C_i$ pertaining to the states $\Psi_i$ for a pair of $^2\Sigma$ molecules
with $\Xi/B\sim10^{-5}$ as a function of an inhomogeneous magnetic field such that $(\eta_m)_1\equiv\eta_m=(\eta_m)_2/1.15$. The concurrence of all four states is unity at $\eta_m=0$, indicating that these are maximally entangled  Bell states.}
\label{fig:Concurrence}
\end{figure}

The entanglement of the two molecules in the various eigenstates of the composite system can be quantified by evaluating their concurrence. To this end, the Hamiltonian of the composite system in the presence of the fields is set up in the composite basis set (i.e., $\{\ket{00}, \ket{01}, \ket{10}, \ket{11}\}$), cf. Fig. \ref{fig:ModelSystem}, which gives rise to a $4\times4$ matrix whose elements are calculated numerically. The eigenproperties of this matrix, obtained by a numerical diagonalization, are all that is needed in order to calculate  the pairwise concurrence from the equations below \citep{Wootters_PairOfQubits, Wootters_EntanglementOfFormation}:
\begin{equation} \rho_{i} = \ket{\Phi_{i}}\bra{\Phi_{i}} \end{equation} \begin{equation} \tilde{\rho_{i}} = (\sigma_{y} \otimes \sigma_{y})\rho_{i}^*(\sigma_{y} \otimes \sigma_{y}) \end{equation} \begin{equation}\mathcal{C}(\rho_{i}) = \max\{0, \sqrt{\Lambda_{1}} - \sqrt{\Lambda_{2}} - \sqrt{\Lambda_{3}} - \sqrt{\Lambda_{4}}\} \label{eqn:concurrence} \end{equation}
Here $\ket{\Psi_{i}}$ are the eigenvectors, $\rho_{i}$  the density matrices, $\rho_{i}^*$ the complex conjugate transpose of the density matrices, $\sigma_{y}$ the Pauli matrix and $\Lambda_{i}$ the eigenvalues (in decreasing order) of the non-Hermitian matrix $\rho\tilde{\rho}$, with $\tilde{\rho}$  the density matrix of the spin-flipped state. The latter density matrix $\tilde{\rho}$ can be readily obtained from a new $4\times4$ Hamiltonian written in the swapped combined eigenstate basis, i.e., interchanging $\ket{00}$ and $\ket{11}$ and $\ket{01}$ and $\ket{10}$. Entanglement itself is a monotonously varying convex function of concurrence (see Eq. 7 in \citep{Trinity1}) that can only take values between $0$ and $1$. Thus, the concurrence is zero for unentangled states and unity for maximally entangled states. 

Here we make the following observation regarding concurrence, which suggests how to tune entanglement by engineering the ``right'' Hamiltonian, a non-trivial inverse problem. Consider a general case of an eigenvector of the $4\times4$ Hamiltonian matrix in the computational basis,
\begin{equation}
\ket{\Phi_{i}}=\begin{pmatrix}
a_i \\
b_i \\
c_i \\
d_i
\end{pmatrix},
\end{equation}
where $a_i,b_i,c_i,d_i$ are the expansion coefficients of the eigenfunction $\Psi_{i}$ in the qubit basis states that fulfil the normalisation  $a_i^2+b_i^2+c_i^2+d_i^2=1$. Then  analytic eigenvalues of the matrix $\rho_i\tilde{\rho}_i$ are  $\{0,0,0,4(b_ic_i-a_id_i)^2\}$. Thus, from Eq. \ref{eqn:concurrence}, the concurrence  corresponding to each qubit  state is given by 
\begin{equation}
C(\rho_i)=\max\{0,2\lvert b_ic_i-a_id_i \rvert \}=2\lvert b_ic_i-a_id_i \rvert 
\end{equation}
The concurrences for the field free case of a pair of molecules with $\Xi/B=10^{-5}$ were found to be all equal to one, thus confirming our earlier observation that the four  states in question were maximally entangled. With an increasing magnetic field, all four states rapidly disentangle. The drop in the concurrence of the two intermediate states is slower than that of the top- and bottom-most states. Furthermore, if the magnetic field is homogeneous, i.e., $\mathcal H_1=\mathcal H_2$, the concurrence for the two intermediate states is found to be exactly equal to one, cf. panel (b) of Fig. \ref{fig:EnergySchematic}. We note that the application of an inhomogeneous magnetic field is akin to effecting a measurement on $M_i$ for both molecules that destroys their entanglement, cf.  panel (c) of Fig. \ref{fig:EnergySchematic}.

\section{CNOT implementation schemes}
\label{Sec: 3}
The conditional quantum dynamics needed to implement a CNOT gate is  provided by a bipartite two-state system with a mutual interaction. Entangled spin systems in NMR \citep{Barenco_Cond_Dynamics}, meanwhile banished as impractical for implementing a quantum computer \cite{NMR_flop}, offer ideas and guidance for the study of isomorphous systems, such as spin $\frac{1}{2}$ molecules entangled by the electric dipole-dipole interaction, considered here. 

In both schemes that we describe below, an initial state of the system is prepared in the presence of an inhomogeneous magnetic field (an electric field may or may not be present) and given as input to the `black-box' that performs the gate operation. This initial state may exist in any one of the four states shown in panel (c) of Fig. \ref{fig:EnergySchematic} or in a superposition. 
\begin{equation}
\ket{\Psi_{\text{input}}}=a\ket{00}+b\ket{01}+c\ket{10}+d\ket{11}
\end{equation}
 A CNOT operation on this initial state results in the following final state:
\begin{equation}
\ket{\Psi_{\text{output}}}=a\ket{00}+b\ket{11}+c\ket{10}+d\ket{01}
\end{equation}
where the state of the second molecule (molecule 2), cf. Fig. \ref{fig:Schematic}, acted as a control qubit and the state of the first molecule (molecule 1) as the target qubit: If the control qubit is ``high'' (i.e., $1$), the target qubit gets flipped (i.e., changes from $0$ to $1$ or from $1$ to $0$), whereas if the control qubit is ``low'' (i.e., $0$), the target qubit remains unaltered.

\subsection{Scheme I}
\begin{figure}
\centering
\includegraphics[width=1\textwidth, height=\textheight, keepaspectratio]{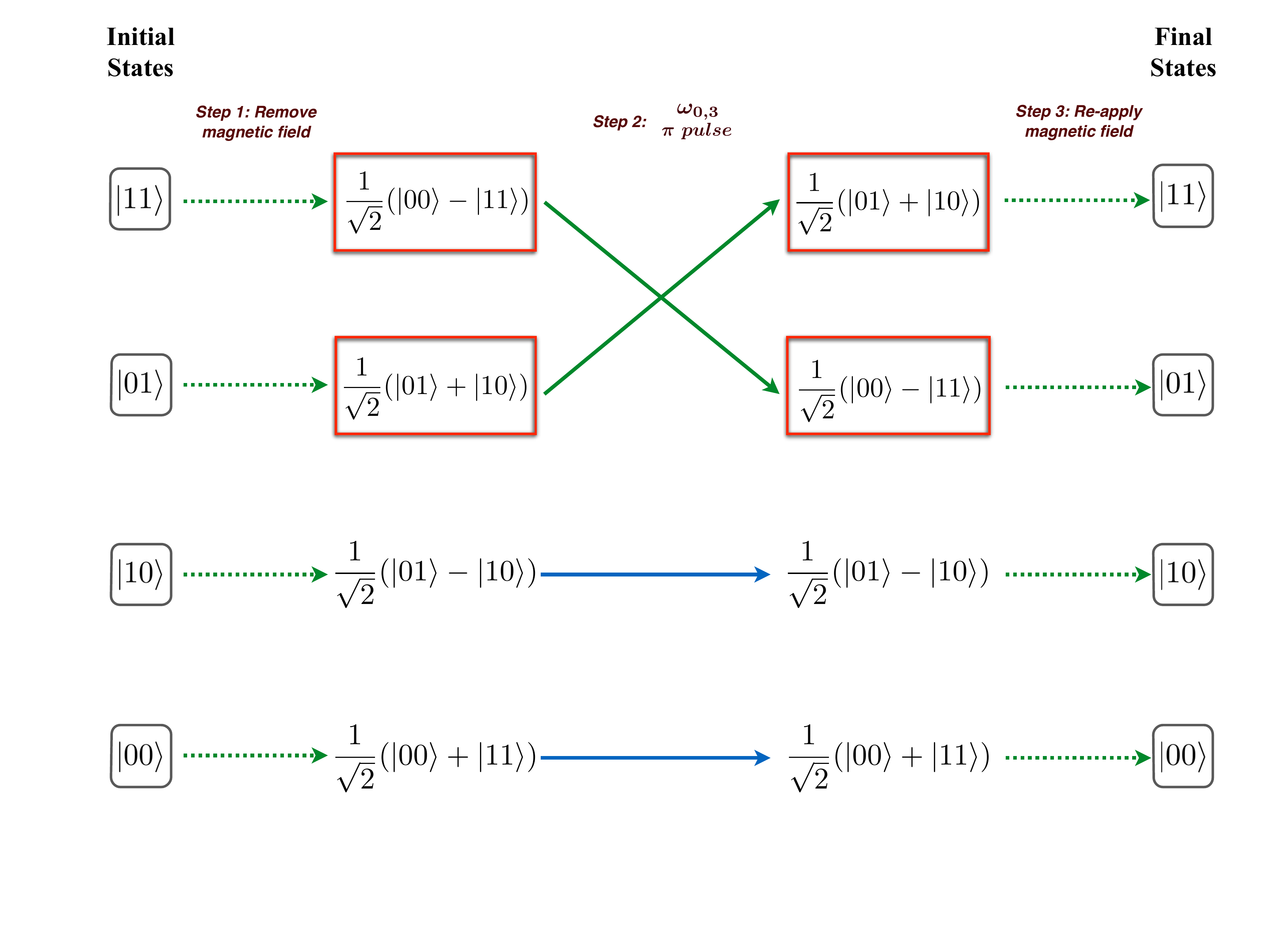}
\caption{Scheme I for CNOT quantum logic gate implementation, with the second qubit used as a control bit. A three step process that involves adiabatically removing and re-applying an inhomogeneous magnetic field. Green arrows depict a change in state due to an operation and the key Bell states involved in the gate operation are boxed in red (cf. also Fig.\ref{fig:EnergySchematic}).}
\label{fig:Scheme1}
\end{figure}

The first scheme that we propose takes direct advantage of the electric dipole-dipole coupling that results, in the absence of a magnetic field, in maximally entangled states. Figure \ref{fig:Scheme1} outlines the three-step process. Beginning with an initial state of the two-molecule system in the presence of  an inhomogeneous magnetic field (an electric field may or may not be present as well), the first step is to adiabatically reduce the magnetic field to zero. We expect the initial state to adiabatically evolve into its corresponding Bell state as shown in Fig. \ref{fig:EnergySchematic}. Next, we apply a $\pi$ pulse resonant with the energy difference between the states $\ket{\Psi_1}=\frac{1}{\sqrt{2}}(\ket{00}-\ket{11})$ and $\ket{\Psi_2}=\frac{1}{\sqrt{2}}(\ket{01}-\ket{10})$, i.e., corresponding to a frequency $\omega_{0, 3}$. This results in interchanging the populations of the top two states while leaving the remaining states unchanged. In the third and final step, the inhomogeneous magnetic field is adiabatically re-introduced, whereby the Bell state is transformed into one of the four decoupled basis states (which form the computational basis). The advantage of this scheme is the high degree of entanglement between the two qubits, available in the field free case.

\subsection{Scheme II}

\begin{figure}
\centering
\includegraphics[width=1\textwidth, height=0.8\textheight, keepaspectratio]{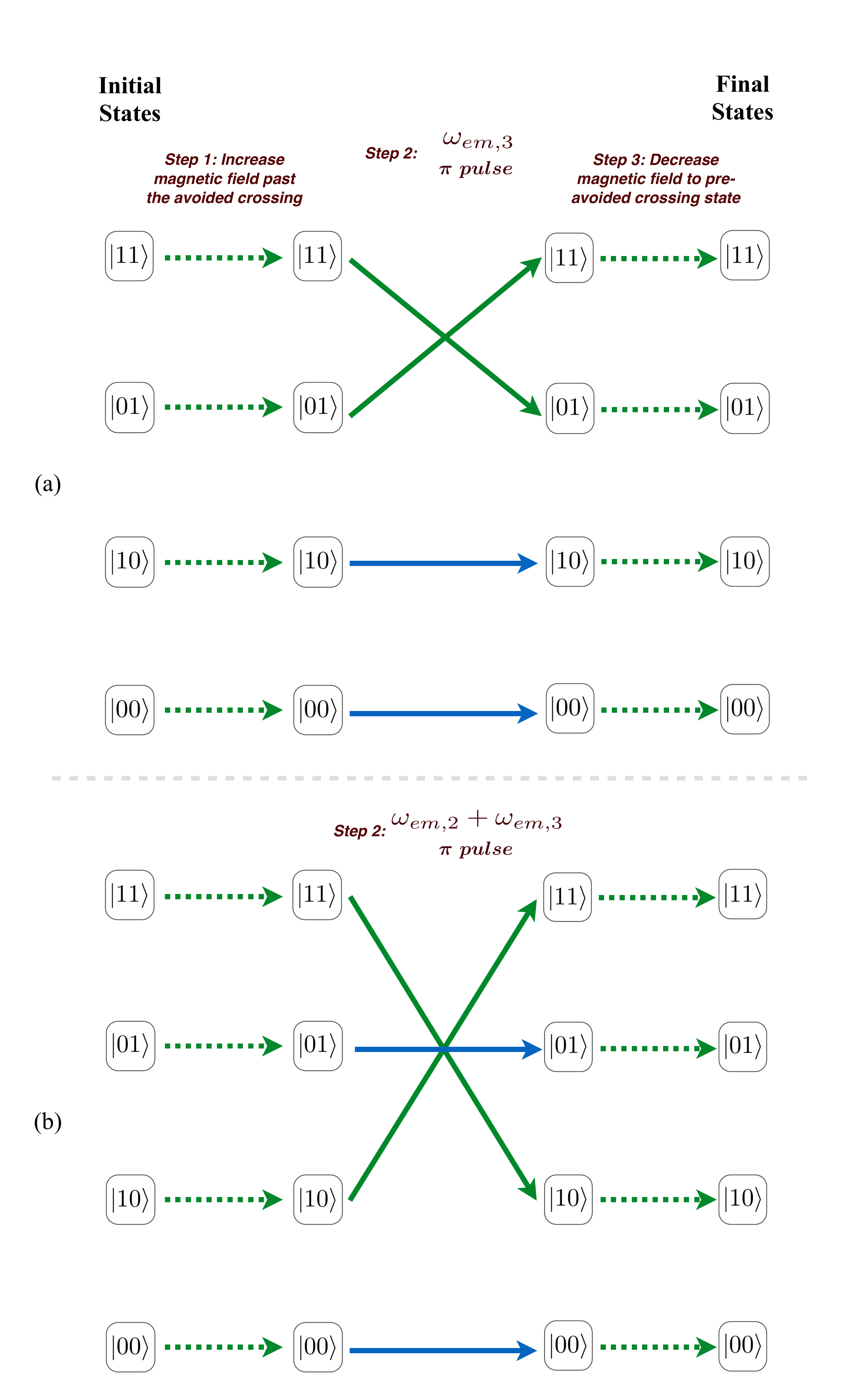}
\caption{Scheme II for CNOT quantum logic gate implementation, with the second (first) molecule used as the control qubit in the top (bottom) panel. Scheme II,  a three-step process,   involves an adiabatic evolution of the system beyond an avoided crossing, applying a $\pi$ pulse of requisite resonance frequency, and  finally devolving the system back to a pre-avoided crossing state for readout of individual qubits.}
\label{fig:Scheme2}
\end{figure}

According to Scheme II, given an initial state of the system, a CNOT operation  consists of three steps. Firstly,  an inhomogeneous magnetic field is adiabatically increased to bring the system beyond an avoided crossing. This step allows to operate the gate under conditions where the key frequencies needed for the system's optical control are well resolved. The next step is to apply a $\pi$ pulse resonant with the desired shift in order to interchange the populations of the $\ket{01}$ and $\ket{11}$ states. Panel (a) in Figure \ref{fig:Scheme2} illustrates the three step process for the case when the second molecule acts as a control qubit. Hence, the frequency of the $\pi$ pulse is $\omega_{em, 3}$. In panel (b), the first molecule is taken to be the control qubit and hence a $\pi$ pulse of frequency $\omega_{em, 2}+\omega_{em, 3}$ is required in order to interchange the populations of the  $\ket{10}$ and $\ket{11}$ states. In the third and final step, the system is adiabatically brought back to a pre-avoided crossing state for final readout of the individual qubits. We note that a read out of the individual qubits beyond the avoided crossing would not be possible because of the mixing with higher states there. 
 
Unlike in Scheme I, in Scheme II the entanglement between the two qubits is low during the entire gate operation. However, entanglement may be  induced dynamically \cite{Trinity1,Trinity3}. As indicated below,  Scheme II scores over Scheme I in terms of practicality.

\section{Results and discussion}
\label{Sec: 4}
In this section we describe the results of our quantitative study of Schemes I and II for CNOT gate implementation using a pair of NaO molecules. The molecular constants of NaO relevant to this study are listed in Table \ref{table:NaO}. We note that for Scheme I to be successfully implemented, the frequencies $\omega_{0, 1}$ and $\omega_{0, 3}$ must be well resolved. This condition is met at high values of the dipole-dipole interaction parameter $\Xi/B$, Eq. (\ref{eqn:Xi}). In the case of NaO, this would require an intermolecular distance that is significantly less than the benchmark value of  $500$ nm given by what is attainable in optical lattices produced by fiber lasers operating at about a $\mu$m (however, cf. recent work in refs. \citep{LiRb,Bomble_NaCs}). In contrast, Scheme II is not as  demanding in regard to the strength of the electric dipole-dipole coupling and so conditions for its implementation are met even at the benchmark intermolecular separation of $500$ nm.  

\begin{table*}[!ht]
\centering
\caption{\small Rotational constant, $B$, spin-rotation constant, $\gamma$, electric dipole moment, $\mu$,  and values of the dimensionless interaction parameters $\eta _{el}$ and $\eta _{m}$ at electric and magnetic fields of 100 kV/cm, 1 Tesla, respectively, for  NaO(A$^2\Sigma$); also shown is the value of the electric dipole-dipole interaction parameter $\Xi$, see Eqs. (\ref{etam})-(\ref{eqn:Xi}). Compilation based on Refs. \cite{Worsnop-Herschbach} and our own calculations. $^a$Calculated using Gaussian 09.  $^b$Becke3LYP type calculation using TZP-DKH basis \cite{feller1996role,schuchardt2007basis}.}
\vspace{.3cm}
\label{table:NaO}
\begin{tabular}{| c | c | c | c | c | c | c | c | c |}
\hline 
\hline
$B$ [cm$^{-1}$] & $\gamma$ [cm$^{-1}$] & $\mu$ [D] &  $\eta_{el}$ @ 100 kV/cm & $\eta_{m}$  @ 1 T & $\Xi$ [cm$^{-1}$] @ 500 nm \\[5pt]

\hline

0.462 & 0.193 & 7.88$^{a,b}$  & 3.63 & 2.02  & $2.49 \times 10^{-6}$ \\[5pt]

 \hline
 \hline
  
\end{tabular}
\end{table*}

\subsection{Scheme II for a pair of NaO molecules}
\begin{figure}
\centering
\includegraphics[width=0.7\textwidth, height=\textheight, keepaspectratio]{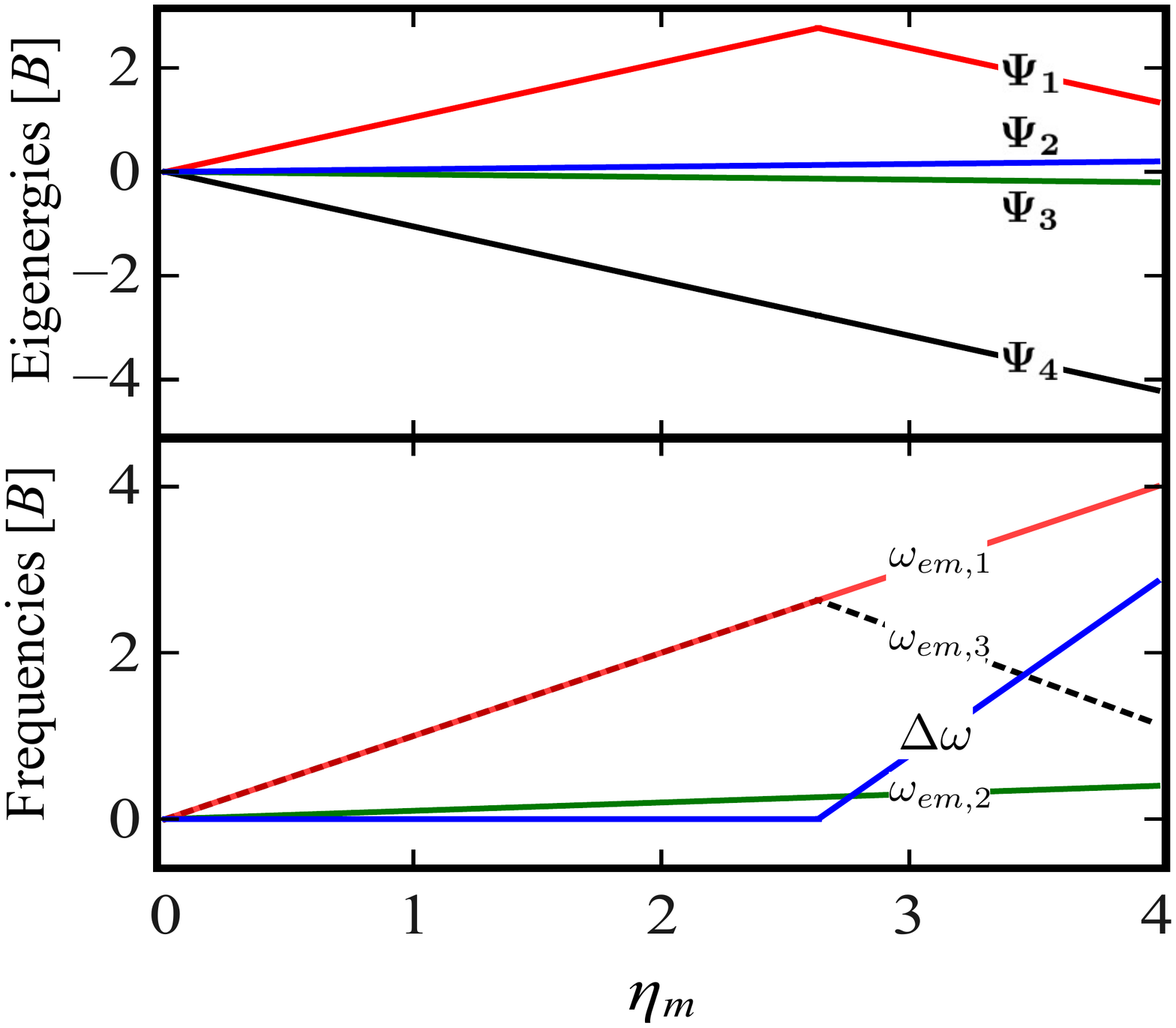}
\caption{In the upper panel, the four eigenenergies of the system for the field parameters $(\eta_{el})_1=(\eta_{el})_2=0$ and $(\eta_m)_2/(\eta_m)_1=1.1$ are shown as functions of $\eta_m$. The three transition frequencies $\omega_{em,1}$, $\omega_{em,2}$, and $\omega_{em,3}$ along with the key frequency difference $\Delta \omega = \omega_{em, 3} - \omega_{em, 1}$ are plotted as functions of $\eta_m$ in the lower panel. Note the change in behaviour of $\Psi_1$ and $\Delta\omega$ at the avoided crossing.}
\label{fig:EnergyAndOmegaPlots}
\end{figure}

Here we present results for a pair of NaO($^2\Sigma$) molecules trapped  in an optical array $500$ nm apart and subject to an inhomogeneous  magnetic field (in the absence of an electric field).  Throughout the operation, an inhomogeneity in the magnetic field is maintained between the two sites such that $(\eta_m)_2/(\eta_m)_1=1.1$. By making use of the procedure outlined in Section \ref{Sec: 2B}, we track the bottom four states of the system adiabatically while increasing $(\eta_m)_1$ and $(\eta_m)_2$. For the case of zero fields, the bottom-most states are  Bell states in which the qubits of our choice are fully entangled. An inhomogeneous magnetic field decouples them and puts the system in any of the composite basis states, \{$\ket{00}$, $\ket{01}$, $\ket{10}$, $\ket{11}$\}. The upper panel of Figure \ref{fig:EnergyAndOmegaPlots} shows the four eigenenergies of these tracked states while the lower panel shows three transition frequencies and the key frequency difference, $\Delta \omega$. It can be seen that an avoided crossing is encountered between $\eta_m=2.63$ and $\eta_m=2.64$, where the highest state changes its character from a low-field seeker to a high-field seeker. The  frequency difference $\Delta \omega$ shows a sharp rise after the avoided crossing (see the blue in the lower panel of Fig. \ref{fig:EnergyAndOmegaPlots}). As a result, at such enhanced  $\Delta \omega$ the frequencies $\omega_{em,1}$ and $\omega_{em,3}$ can be resolved. Hence, a CNOT operation according to Scheme II would involve the application of a pulse resonant with the frequency $\omega_{em,3}$, with  molecule 2 acting as a control qubit. Table \ref{Scheme2_Avoided_Xing_table} lists the three key frequencies as well as the frequency $\Delta\omega$ before and after the avoided crossing. For a readout of the individual qubits, the system has to be  adiabatically devolved to a state before the avoided crossing (at, say, $\eta_m=2.63$). In Table \ref{IndividualMolecule_Table}, we list the frequencies needed to flip the individual molecules between their $\ket{0}$ and $\ket{1}$ states. Furthermore, we find that the contribution to the frequencies due to the dipole-dipole coupling term is very small.  Consequently,  diagonalising the Hamiltonians of the individual molecules separately and using the eigenenergies thus obtained to cast the system Hamiltonian in the composite basis was a very good approximation to make in Refs. \citep{Trinity1, Trinity2}). Thus
\begin{eqnarray}
\braket{\Psi_{initial}|H_{System}|\Psi_{final}}  \approx  
\braket{\Psi_{initial}|H_{1}|\Psi_{final}} + \braket{\Psi_{initial}|H_{2}|\Psi_{final}} \nonumber \\ + \braket{\Psi_{initial}|V_{d-d}|\Psi_{final}} 
\end{eqnarray}
where $\Psi_{initial}$ and $\Psi_{final}$ refer to the eigenstates of the combined two molecule system. The matrix elements $\braket{\Psi_{initial}|V_{d-d}|\Psi_{final}}$ comes out to be six orders of magnitude smaller than the other two terms. As a result, the weak dipole-dipole coupling and the small entanglement in the presence of an inhomogeneous field are actually responsible for helping us achieve individual qubit addressability.

\begin{table}[]
\centering
\caption{Exploiting the occurrence of an avoided crossing in Scheme II}
\label{Scheme2_Avoided_Xing_table}

\begin{tabularx}{0.75\textwidth}{c*{4}{|Y}|}
\cline{2-5}
 & \multicolumn{2}{c|}{Before avoided crossing} & \multicolumn{2}{c|}{After avoided crossing} \\ \hline 
\multicolumn{1}{|c|}{}& Molecule 1 & Molecule 2 & Molecule 1 & Molecule 2 \\ \cline{2-5}
\multicolumn{1}{|c|}{$\eta_m$} & 2.63 & 2.893 & 2.64 & 2.904 \\
\multicolumn{1}{|c|}{$H$} & 1.302 T & 1.432 T & 1.307 T & 1.438 T \\ \cline{2-5}
\multicolumn{1}{|c|}{$\omega_{em, 1}$} & \multicolumn{2}{c|}{36.427 GHz} & \multicolumn{2}{c|}{36.427 GHz}  \\
\multicolumn{1}{|c|}{$\omega_{em, 2}$} & \multicolumn{2}{c|}{3.643 GHz} & \multicolumn{2}{c|}{3.656 GHz} \\
\multicolumn{1}{|c|}{$\omega_{em, 3}$} & \multicolumn{2}{c|}{36.427 GHz} & \multicolumn{2}{c|}{36.368 GHz} \\
\multicolumn{1}{|c|}{$\Delta\omega = \omega_{em, 1} - \omega_{em, 3}$} & \multicolumn{2}{c|}{1.662 Hz} & \multicolumn{2}{c|}{198.429 MHz}  \\ \hline
\end{tabularx}
\end{table}

\begin{table}[]
\centering
\caption{Addressing individual molecules for readout in Scheme II. Here $\Delta E$ is the energy difference between the eigenenergies of $\ket{0}$ and $\ket{1}$ states of the individual  molecules.}
\label{IndividualMolecule_Table}

\begin{tabularx}{0.75\textwidth}{c*{4}{|Y}|}
\cline{2-3}
 & \multicolumn{2}{c|}{Before avoided crossing}  \\ \hline 
\multicolumn{1}{|c|}{}& Molecule 1 & Molecule 2 \\ \cline{2-3}
\multicolumn{1}{|c|}{$\eta_m$} & 2.63 & 2.893 \\ 
\multicolumn{1}{|c|}{$H$} & 1.302 T & 1.432 T \\ \cline{2-3}
\multicolumn{1}{|c|}{$\Delta E$} & 36.427 GHz & 40.069 GHz \\ \cline{2-3}
\hline 
\end{tabularx}
\end{table}

\subsection{Feasibility of Schemes I and II}
\subsubsection{Broadening}
\begin{table}[]
\centering
\caption{Broadening of the three key frequencies when addressing the composite two-molecule system at $\eta_m$=2.64, where $i\in \{1, 2, 3\}$. See Fig. \ref{fig:Broadening}.}
\label{Broadening_together}
\begin{tabular}{l|l}
Quantity & Broadening (=$\omega_{ai}$ - $\omega_{bi}$) \\ \hline \hline
$\omega_{em, 1}$ &  0.069 Hz\\
$\omega_{em, 2}$ & 0.013 Hz \\
$\omega_{em, 3}$ & 2.461 kHz \\ 
\end{tabular}
\end{table}

\begin{figure}
\centering
\includegraphics[width=0.9\textwidth, height=\textheight, keepaspectratio]{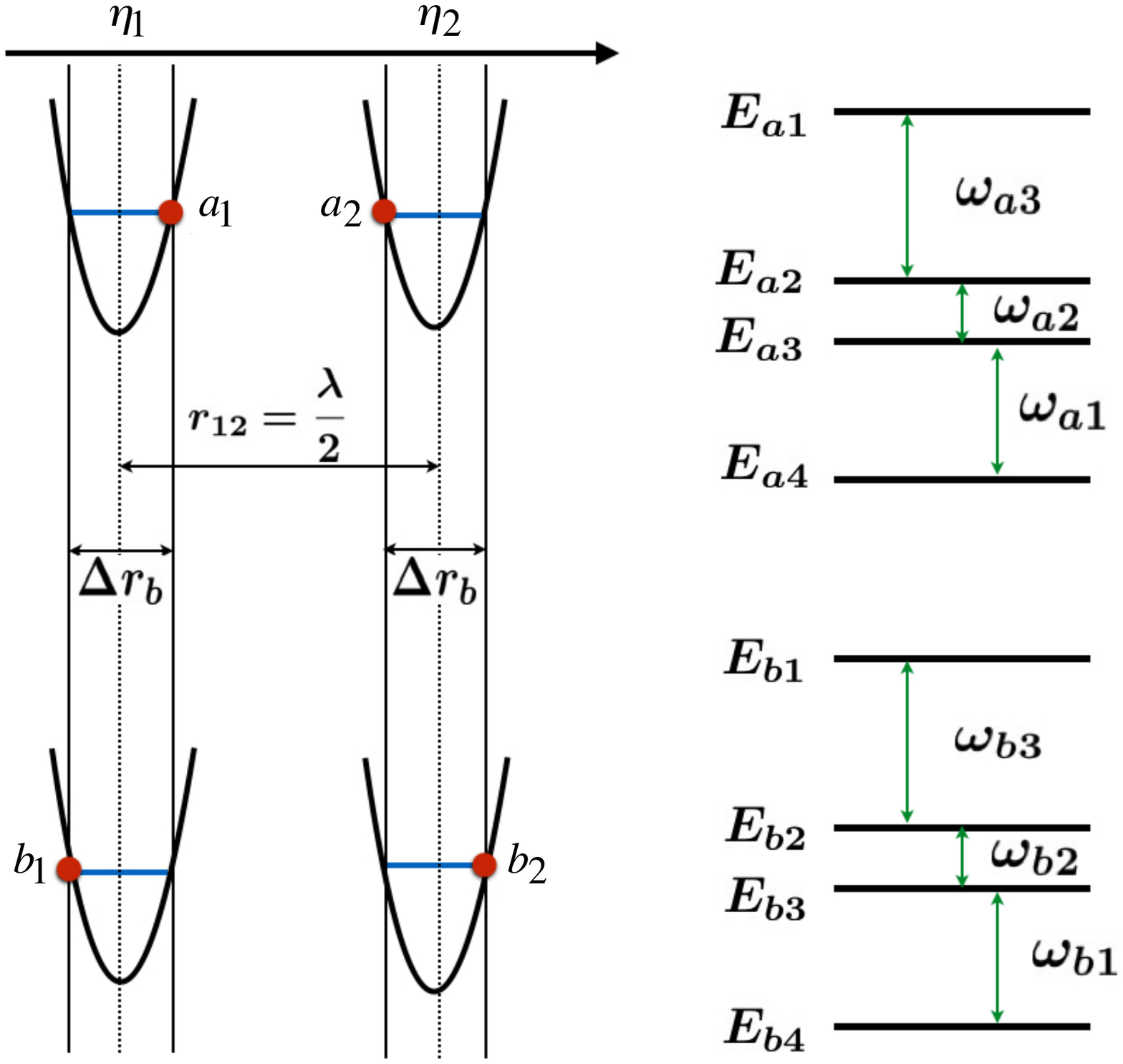}
\caption{Maximum possible broadening of spectral lines when addressing the composite two-molecule system. See text.}
\label{fig:Broadening}
\end{figure}

\begin{figure}
\centering
\includegraphics[width=1\textwidth, height=\textheight, keepaspectratio]{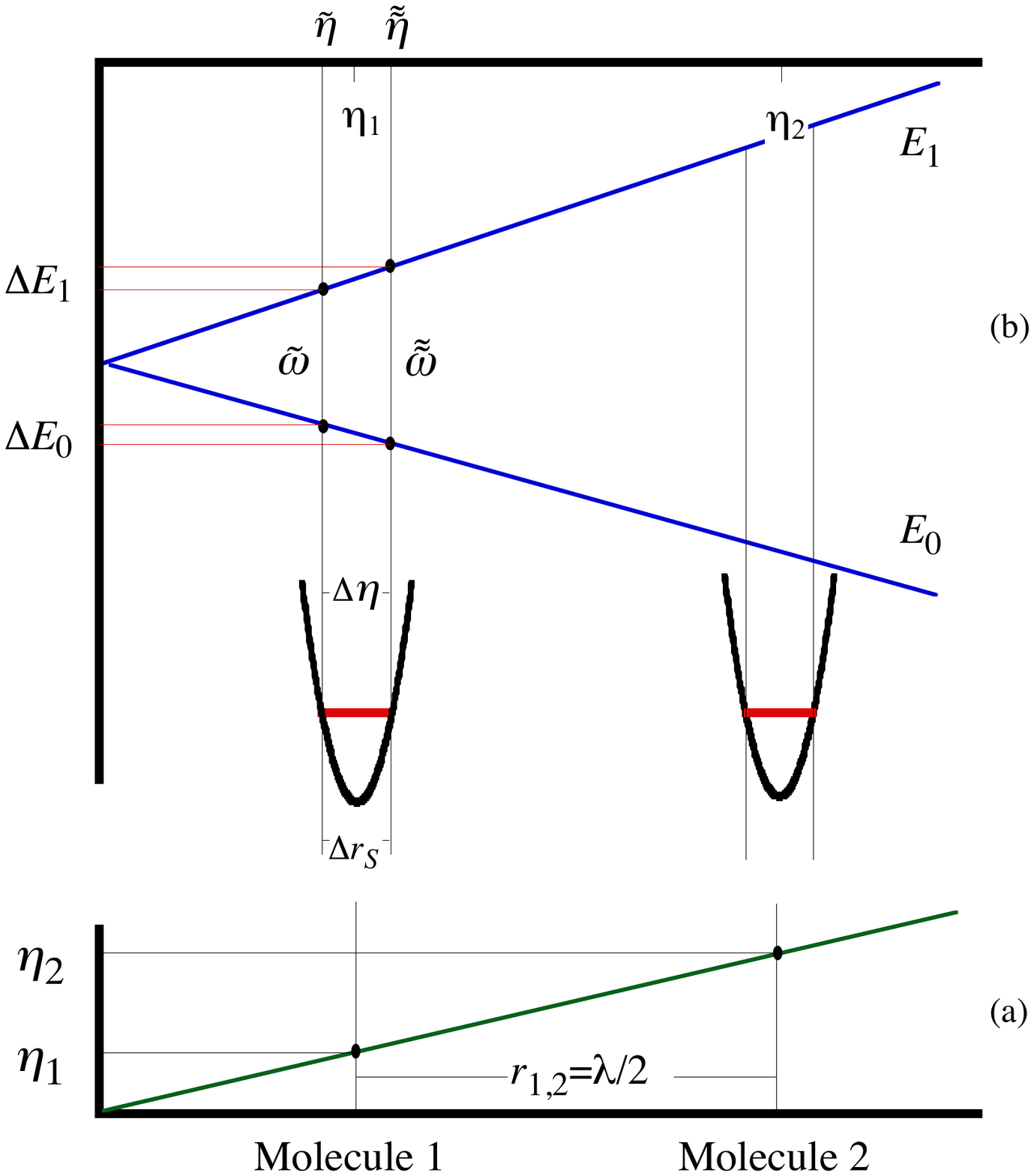}
\caption{A schematic view of the maximum possible broadening of spectral lines, $\tilde{\tilde{\omega}}-\tilde{\omega}$, for each individual molecule. Panel (a) shows the broadening for transitions between high- and low-field seeking states. Panel (b) shows the linear dependence of the applied field strength on the longitudinal coordinate, r, along the array. Adjacent wells confining molecule 1 and molecule 2 are separated by a distance $r_{1,2}=\lambda/2$, with $\lambda$ the wavelength of the optical trapping field. In this figure, the interaction parameter $\eta$ stands for both $\eta_m$ and $\eta_{el}$.}
\label{fig:Broadening_individual}
\end{figure}

An important issue regarding the feasibility of the proposed schemes is the broadening of the spectral lines of the system. Figures \ref{fig:Broadening} and \ref{fig:Broadening_individual} illustrate the broadening of spectral lines when treating the two molecules as a composite system or individually. In the former case, the broadening arises due to the dipole-dipole coupling between the molecules and the (linear) inhomogeneity of the external field(s). The spread in the translational confinement of each molecule over the range $\Delta r_{b}$ in the trap gives rise to two extreme cases as illustrated in Fig. \ref{fig:Broadening}. Case (a) pertains to the minimum and case (b) to the maximum possible value of $r_{1,2}$. Recall that the intermolecular separation $r_{1,2}$ directly influences the frequencies $\omega_{1}$, $\omega_{2}$ and $\omega_{3}$, cf. panel (c) of Fig. \ref{fig:EnergySchematic} (the smaller $r_{1,2}$, the greater the dipole-dipole coupling element, and hence the greater the energy differences between the four states). For determining the maximum possible broadening, it is sufficient to consider the minimum and maximum values of $r_{1,2}$ and compare the corresponding broadening with the key frequency difference $\Delta \omega = \omega_{em, 3} - \omega_{em, 1}$. In each of the two extreme cases, the system can occupy one of the four possible states, with eigenenergy $E_{ij}$, for $i = \{a, b\}$ and $j = \{1, 2, 3, 4\}$. For any given transition between two of these four states, the maximum possible broadening due to the spread in the translational confinement of the molecules will be the difference of corresponding frequencies in cases (a) and (b). Thus, the condition for feasibility of CNOT gate operation in Scheme II is that there be no overlap of these frequency ranges,
\begin{equation} 
[\omega_{a3}, \omega_{b3}] \cap [\omega_{a2}, \omega_{b2}] \cap [\omega_{a1}, \omega_{b1}] \in \{\emptyset\}, 
\end{equation}
where $\omega_{i1}$, $\omega_{i2}$ and $\omega_{i3}$ denote the differences $E_{i1}-E_{i2}$, $E_{i2}-E_{i3}$ and $E_{i3}-E_{i4}$ respectively, for $i=\{a, b\}$. For the microkelvin optical trap conditions envisaged in our setup \citep{DeMille, grimm_et_al, del_r_b_2_Wei_Xue, Bloch_ManyBodyGases}, we take $\Delta r_{b} = 30 \ nm$, and find that the broadening due to the dipole-dipole coupling is about six orders of magnitude greater than that due to the inhomogeneity of the fields. Broadening values of the three key frequencies are listed in Table \ref{Broadening_together}. Note that the broadening of $\omega_{em, 3}$ is five orders of magnitude smaller than that of the key frequency shift $\Delta \omega$ at $\eta_m=2.64$.

 In the latter case of addressing both molecules individually, broadening can be defined as the difference of the `flipping frequencies' of the molecule at the two extremes of the optical trap, see Fig. \ref{fig:Broadening_individual}. Thus, the feasibility criteria for the individual addressability of the qubits can be expressed in the form of the following inequalities for the two molecules:
\begin{equation}
\Delta\omega_{1} = |\tilde{\omega_1}-\tilde{\tilde{\omega_1}}| \ll |\Delta E_{0}(\eta_1)-\Delta E_{1}(\eta_2)| 
\end{equation}
\begin{equation}
\Delta\omega_{2} = |\tilde{\omega_2}-\tilde{\tilde{\omega_2}}| \ll |\Delta E_{0}(\eta_1)-\Delta E_{1}(\eta_2)|
\end{equation}
where the indices 1 and 2 refer to the two molecules (also see Table \ref{IndividualMolecule_Table}, 3rd row); i.e. the broadening for each molecule must be very small compared to the difference of the flipping frequencies of the two molecules. Both of the above conditions are met by our candidate system of a pair of NaO molecules. 
 


 
\section{Conclusions}
\label{Sec: 5}
We have examined the eigenproperties of a pair of $^2\Sigma$ molecules in the presence of superimposed electric and magnetic fields and proposed two  schemes for the implementation of the controlled-NOT quantum gate. Preliminary results for a pair of NaO molecules show a non-zero transition dipole moment corresponding to the transition frequency $\omega_{em, 3}$ and a feasibility of an optical control in the face of the line broadening arising from the dipole-dipole interaction and the inhomogeneity of the fields. Our choice of qubits is consistent with the possibility of implementing  field-free two qubit Bell states or multi-qubit highly entangled states (cluster states, GHZ states or W states) in two-dimensional and three-dimensional arrays for one-way quantum computation \citep{BriegelRaussendorf_Entanglement1, One-way_QC, BriegelRaussendorf3}. In our forthcoming work the schemes proposed will be  tested by invoking multi-target optimal control theory (MTOCT) \citep{Riedle_MTOCT1, Riedle_MTOCT2, Trinity3,QuantumOptimalControl} as a means of optimizing the initial-to-target transition probability via a tailored optical control field and evaluating the attainable fidelity. 
 
We note that the second step of both schemes could be replaced by the more robust Adiabatic Population Transfer process \citep{APTNeutralMolecules}, whereby all three steps would be rendered adiabatic. Furthermore, we note that by choosing the highest $\tilde{N}=0$ and the lowest $\tilde{N}=1$ states as the $\ket{0}$ and $\ket{1}$ qubits, respectively, the avoided crossing between them could be used for adiabatic quantum computation, as in Ref. \citep{AQC_Cooper_Pairs}, or even for holonomic quantum computation \citep{HolonomicQC_AvoidedCrossings, Zanardi_HQC}.

\begin{acknowledgments}
Support by the DFG through grant FR 3319/3-1 is gratefully acknowledged. MK thanks the International Max Planck Research School (IMPRS) for a stipend.
\end{acknowledgments}

\bibliography{QC_References}
\end{document}